\documentclass[a4paper,fleqn,usenatbib]{mnras}

\usepackage{newtxtext,newtxmath}

\usepackage[T1]{fontenc}
\usepackage{ae,aecompl}

\usepackage{graphicx}	
\usepackage{amsmath}	
\usepackage{amssymb}	

\title[Tight stellar binaries in the Orion Nebula Cluster]{Is stellar multiplicity universal? Tight stellar binaries in the Orion Nebula Cluster}

\author[G. Duch\^ene et al.]{
Gaspard Duch\^ene,$^{1,2}$\thanks{The observations presented here were obtained as part of ESO program ID: 096.C-0270.}\thanks{E-mail: gduchene@berkeley.edu}
S. Lacour$^{3}$, E. Moraux$^{2}$, S. Goodwin$^{4}$, J. Bouvier$^{2}$
\\
$^{1}$Astronomy Department, University of California Berkeley, CA 94720, USA\\
$^{2}$Universit\'e Grenoble Alpes, CNRS, IPAG, 38000 Grenoble, France\\
$^{3}$LESIA/Observatoire de Paris, PSL, CNRS, UPMC, Universit\'e Paris Diderot, 5 place Jules Janssen, 92195 Meudon, France\\
$^{4}$Department of Physics and Astronomy, University of Sheffield, Hounsfield Road, Sheffield S3 7RH, UK\\
}

\date{Accepted XXX. Received YYY; in original form ZZZ}

\pubyear{2018}

\begin{document}
\label{firstpage}
\pagerange{\pageref{firstpage}--\pageref{lastpage}}
\maketitle

\begin{abstract}
We present a survey for the tightest visual binaries among 0.3--2\,$M_\odot$ members the Orion Nebula Cluster (ONC). Among 42 targets, we discovered 13 new 0\farcs025--0\farcs15 companions. Accounting for the Branch bias, we find a companion star fraction (CSF) in the 10--60\,au range of 21$^{+8}_{-5}$\%, consistent with that observed in other star-forming regions (SFRs) and twice as high as among field stars; this excess is found with a high level of confidence. Since our sample is dominated by disk-bearing targets, this indicates that disk disruption by close binaries is inefficient, or has not yet taken place, in the ONC. The resulting separation distribution in the ONC drops sharply outside 60\,au. These findings are consistent with a scenario in which the initial multiplicity properties, set by the star formation process itself, are identical in the ONC and in other SFRs and subsequently altered by the cluster's dynamical evolution. This implies that the fragmentation process does not depend on the global properties of a molecular cloud, but on the local properties of prestellar cores, and that the latter are self-regulated to be nearly identical in a wide range of environments. These results, however, raise anew the question of the origin of field stars as the tight binaries we have discovered will not be destroyed as the ONC dissolves into the galactic field. It thus appears that most field stars formed in regions that differ from well-studied SFRs in the Solar neighborhood, possibly due to changes in core fragmentation on Gyr timescales.
\end{abstract}

\begin{keywords}
binaries: visual -- stars: pre-main-sequence -- open cluster and associations: individual: Orion Nebula Cluster
\end{keywords}




\section{Introduction}

The ubiquity of stellar multiplicity in the youngest stellar populations has been long established, proving that this is an inherent feature of the star formation process itself \citep[][and references therein]{duc13}. In order to constrain the mechanism through which multiple systems form, searches have been conducted to identify trends in multiplicity properties besides the strong dependency on primary stellar mass which is generally well reproduced by a wide range of models \citep[e.g.,][]{bat12,del04,goo04,moe10}. 

Much like studies of the initial mass function, one focus has been on the hunt for significant differences between the multiplicity properties of different stellar populations. From the earliest studies of populations of T\,Tauri stars, it was clear that visual binaries are twice as common in nearby SFRs  as they are among field stars of similar masses at separations ranging from tens to thousands of au \citep[][and references therein]{duc99}. However, this high occurrence of visual companions is not universal, as it was later found that stellar populations in young clusters are characterized by a field-like multiplicity rate. This was reported both for open clusters \citep[e.g.,][]{bou97,pat98} and young clusters still associated with their parent molecular cloud \citep[e.g.,][]{pad97,pet98,duc99b}. This is especially true in the ONC which has been targeted by several multiplicity surveys of increasing resolution, scale and sensitivity \citep{pet98,koe06,rei07,kou16} and is the focus of the present study. There are virtually no binary systems in the ONC whose semi-major axis is larger than 1000\,au \citep{sca99}. 

This dichotomy of multiplicity frequency (field-like in stellar clusters, much higher in loose young associations) can be explained by two distinct scenarios; essentially this is a case of nature versus nurture. In one scenario, dense clusters simply form a much reduced number of wide systems due to intrinsic differences in how star formation proceeds in these environments, while the CSF (defined as the ratio of the number of companions to the number of targets) in loose associations approaches 100\%. The alternative is that all SFRs actually form binary systems with essentially universal characteristics but that are subsequently significantly altered by dynamical processes, such as intra-cluster encounters and decay of unstable high-order multiple systems \citep[][and references therein]{goo07}. Given the observations of loose associations, the initial conditions for multiplicity include a rate of visual binaries that is twice as high as that of field stars but many of the wider pairs could be susceptible to destruction in three-body interactions. 

The debate between these two scenarios has been ongoing for over two decades. In short, it is reasonable to assume that the physics of star formation should differ in environments that lead to such different outcomes as a rich stellar cluster and a loose association \citep[e.g.,][]{ste03,goo04}. However, disruption of wide binaries in dense clusters, if they actually form, is inescapable and can occur on very short timescale  \citep[$\lesssim$1\,Myr; see e.g.,][]{kro95}. Interestingly, the multiplicity properties of diverse environments such as the Taurus association and dense clusters such as the ONC and the Pleiades can be reproduced by assuming a universal set of multiplicity properties and allowing internal cluster dynamics to destroy some systems \citep{kro99,kro01,kro03}. Whether this is the correct explanation, however, has been questioned by different groups \citep{kin12,mar14,par14}. Different assumptions about the current and past dynamical states of stellar populations are at the heart of the ongoing debate, but these cannot be easily tested with current observations, which explains why the problem has been lingering for two decades. 

Besides the implications for the star formation process, determining whether or not multiplicity properties are universal at birth has important ramifications for the topic of the origin of field stars. Indeed, while the population of field stars represents a mix of all modes of star formation in the Galaxy, the excess of visual companions among loose associations readily indicates that such SFRs cannot produce the majority of field stars. Under the universal multiplicity properties scenario outlined above, it is in principle possible to infer the typical stellar density of clusters that produce the majority of field stars in an inverse population synthesis approach \citep{kro95,mar11} although, once again, uncertainties about the early dynamical evolution of clusters raise significant uncertainties \citep{par14}.

As discussed above, the ONC has been one of the key stellar populations in solving this puzzle. However, its large distance \citep[388$\pm$5\,pc,][]{kou17} compared to other nearby SFRs (125--140\,pc) has limited the projected separation range probed by past multiplicity surveys to $\gtrsim60$\,au ($\gtrsim0$\farcs15). Most binaries at these large separations are liable to destruction within the first few Myr of the cluster's evolution but, given our current understanding of the past history of the ONC, tighter binaries should be sufficiently tightly bound so as not to be severely affected \citep{kro99,par09}. In other words, the multiplicity properties of systems tighter than 60\,au should be pristine even in the ONC. This enables an immediate test of the universality hypothesis, since under that scenario, one would expect to find the same companion fraction in the ONC as in other SFRs. That fraction would be roughly twice as high as that of field stars, as indicated by observations in various, non-clustered SFRs \citep{kin12}. Measuring the CSF over the same separation range in the ONC is the goal of the present study.

The fundamental limit of past multiplicity studies of the ONC was angular resolution, which was set by the diffraction limit of the instruments in use. Searches with both the Hubble Space Telescope at visible wavelength and large ground-based telescope in the near infrared are limited to companions outside of 0\farcs1--0\farcs15 in order to be sensitive to stellar companions of all masses and not just to equal-mass binaries (and even then, only if the separation exceeds $\lambda / D$, where $\lambda$ is the observing wavelength and $D$ the telescope diameter). In this study, we take advantage of the aperture masking technique to reach the highest resolution on monolithic telescopes and find tighter companions than previous studies. By virtue of the simplicity of the signal introduced by a binary in this interferometric observations, it is possible to detect and characterize companions down to separations of $\lambda / 2D$, or about 0\farcs025 at 2\,$\mu$m on an 8m telescope \citep[see, e.g.,][]{lac11}. At the distance of the ONC, it is therefore possible to detect companions down to projected separations as small as 10\,au. This same technique has been used in the past to probe stellar companions down to 2--5\,au in several nearby SFRs \citep[e.g,][]{kra11,che15}.

The outline of this paper is as follows: we present the sample selection, observations and data reduction in Section\,\ref{sec:setup}, present the results of our survey in Section\,\ref{sec:res}, and discuss them in Section\,\ref{sec:disc}. 


\section{Sample and Observations}
\label{sec:setup}


\subsection{Sample Selection}
\label{subsec:sample}

\begin{figure*}
	\includegraphics[width=0.49\textwidth]{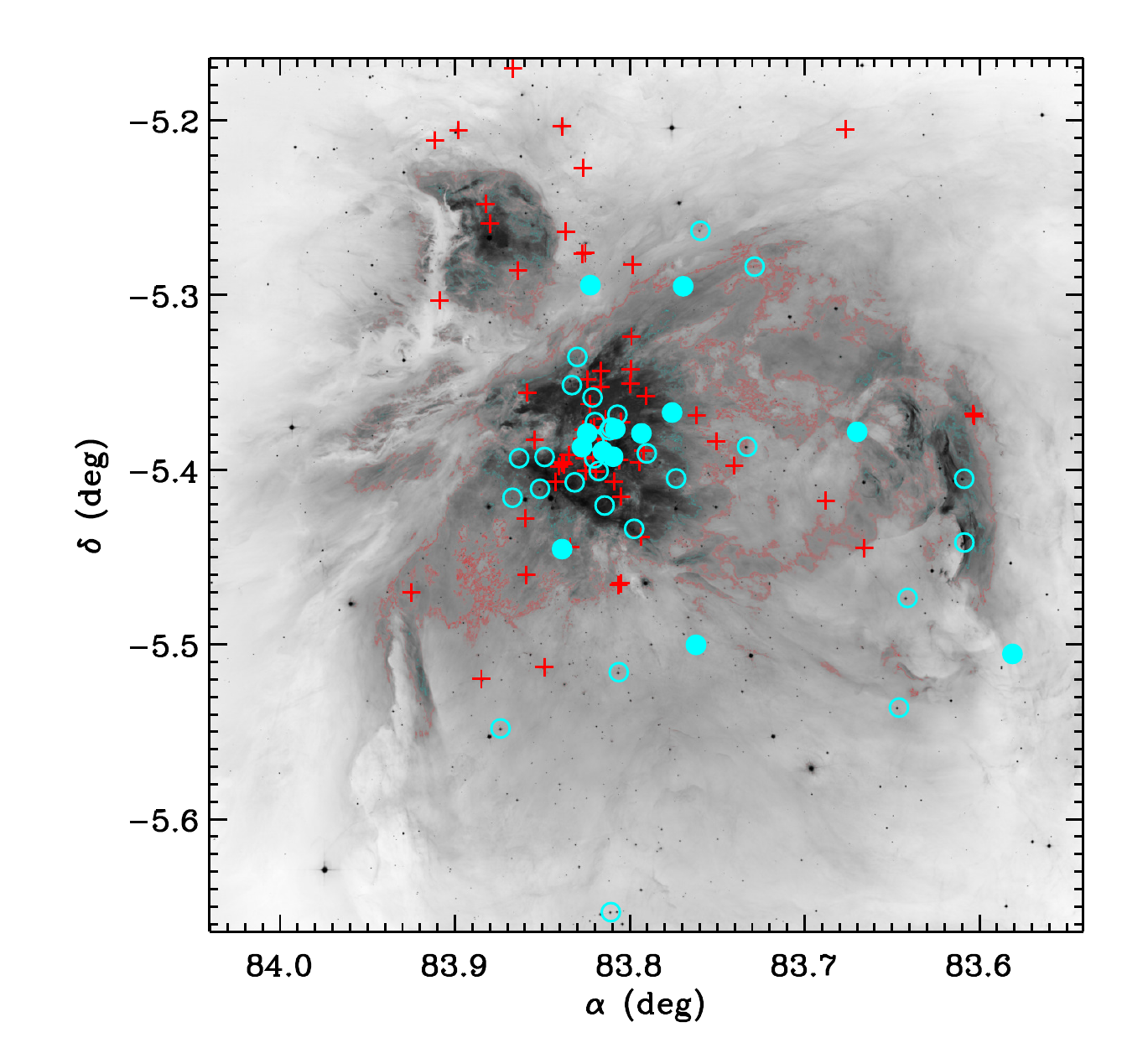}
	\includegraphics[width=0.49\textwidth]{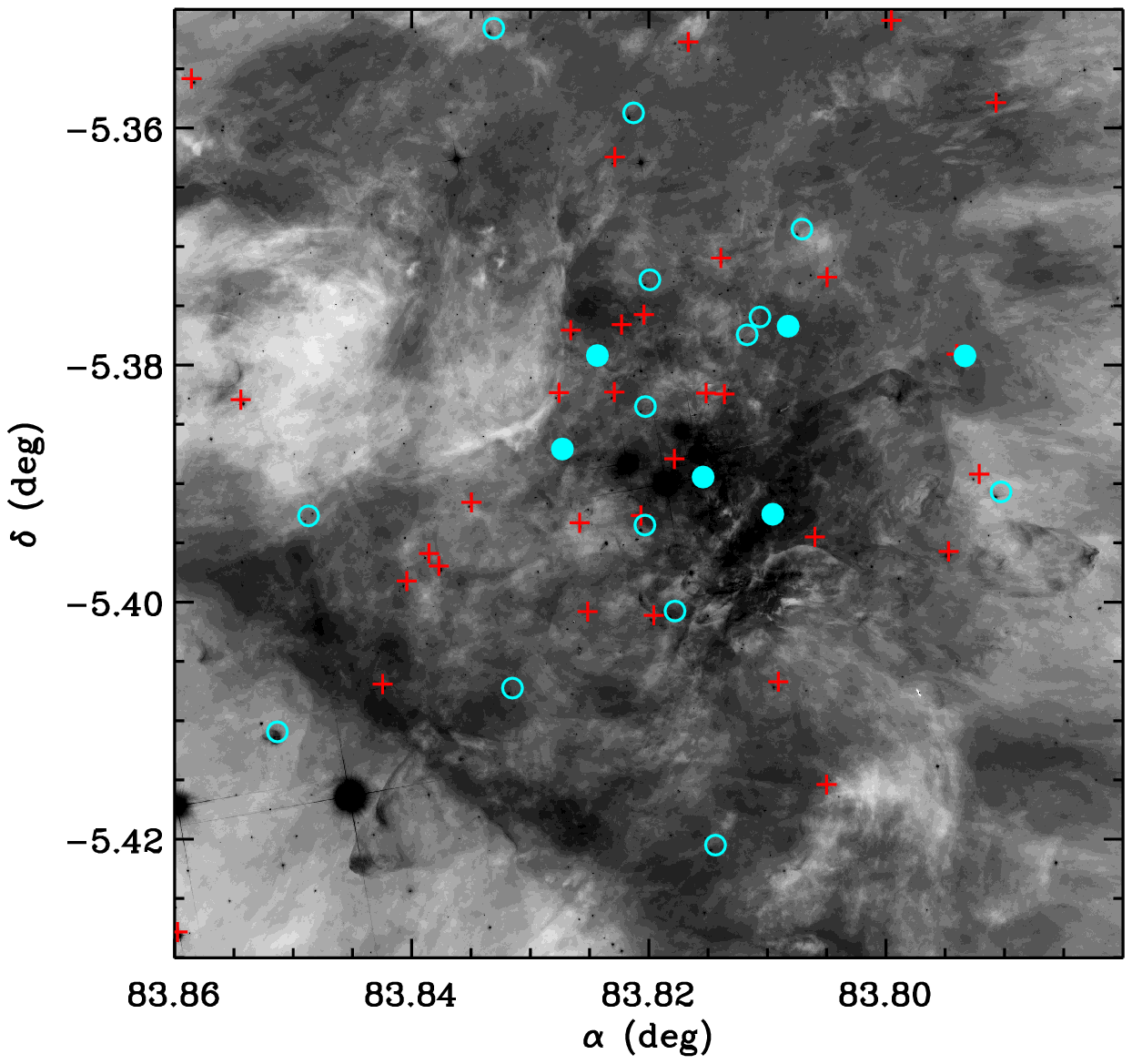}
    \caption{Spatial distribution of the observed (circles) and unobserved (red crosses) subsamples. Filled (open) circles indicate objects that were found to have one (no) companion in the separation range probed by the aperture masking observations (i.e., separation $\leq$0\farcs2). The right panel is a zoom on the center of the cluster. In both cases, the underlying grayscale image is the HST $r$-band image from \citet{rob13}.}
    \label{fig:map}
\end{figure*}

\begin{figure}
	\includegraphics[width=\columnwidth]{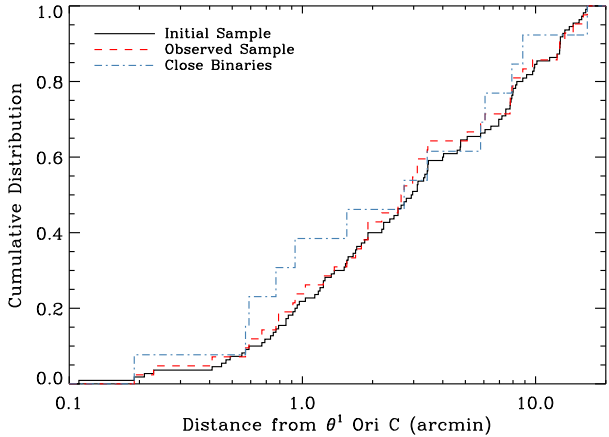}
    \caption{Cumulative distribution of distance to $\theta^1$\,Ori\,C for the initial sample (black solid histogram), the observed subsample (red dashed histogram) and the subset of all binaries with separation $\leq$0\farcs2 (blue dot-dashed histogram).}
    \label{fig:distance}
\end{figure}

\begin{figure}
	\includegraphics[width=\columnwidth]{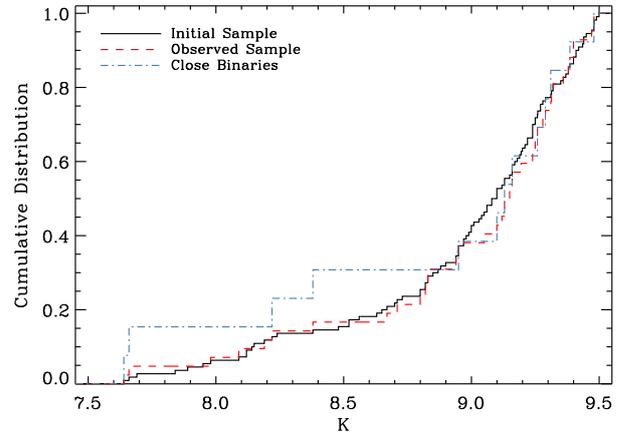}
    \caption{Cumulative $K$-band brightness distribution of the initial sample, the observed subsample and the subset of all binaries with separation $\leq$0\farcs2. Linestyles and colors are as in Figure\,\ref{fig:distance}.}
    \label{fig:kmag}
\end{figure}

The sample was built from the ONC catalogs of \citet[][hereafter H97]{hil97}, \cite{hil98} and \cite{hil13}. From all objects in these catalogs, we first selected a magnitude-limited sample using the range $7.5 \leq K \leq 9.5$. The faint limit is set by a signal-to-noise requirement for successful aperture masking measurement given short exposure integrations. The bright end was chosen to avoid high-mass stars and to ensure that each target would have at least two other targets of similar magnitude that can serve as calibrators. From this list, we discarded objects with spectral types earlier than G0, again to remove stars more massive than $\approx 2 M_\odot$. Finally, objects whose membership probability is less than 50\% \citep{hil97,bou14} were eliminated. At this stage, we retained objects with unknown probability as likely members until proven otherwise; two of those (H97\,3109 and H97\,3131) were subsequently confirmed as cloud members by \cite{fur08}. This yielded our initial sample of 109 targets distributed throughout the ONC, with distances from $\theta^1$\,Ori\,C ranging from 7\arcsec\ to 17\arcmin, i.e., about 2\,pc (see Figure\,\ref{fig:map}). From the initial sample, we observed 42 targets with NaCo-SAM, as well as 4 objects with membership probability lower than 50\%, which we report here for completeness but do not include in our analysis. The basic properties of all observed targets are listed in Table.\,\ref{tab:sample}. A few targets were known subarcsecond binaries and/or spectroscopic binaries from past surveys \citep{tob09, rob13}, although we note that none of these companions could be detected in our aperture masking survey. Figures\,\ref{fig:map} and \ref{fig:distance} illustrate the spatial distribution of the initial and observed samples, while Figure\,\ref{fig:kmag} presents the $K$ band brightness distribution of these samples. 

\begin{figure}
    \includegraphics[width=\columnwidth]{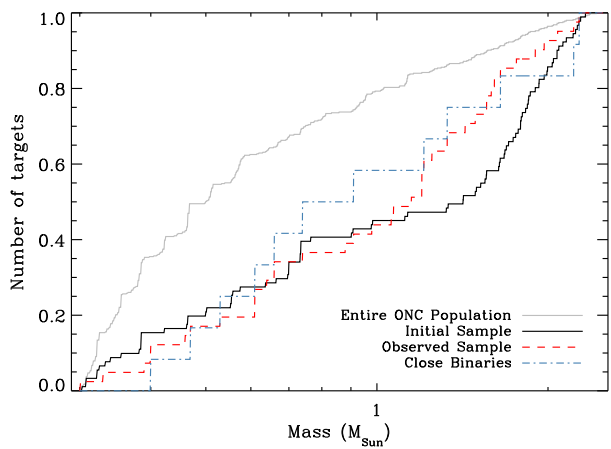}
    \caption{Cumulative Distribution of masses for the initial sample, the observed subsample and the subset of all binaries with separation $\leq$0\farcs2. Linestyles and colors are as in Figure\,\ref{fig:distance}. The mass distribution for the entire ONC is shown in gray for reference, based on the survey by \citet{dar10}.}
    \label{fig:masses}
\end{figure}

Estimating masses in the ONC population is a notoriously non-trivial issue because of crowding, confusion with the surrounding nebula and large and inhomogeneous line of sight extinction. As a result, while many stars in our initial sample have multiple mass estimates in the literature \cite[e.g.,][]{hil97,dar10,man12,dar16,kim16}, there are differences up to a factor of 3 between the various estimates. Stellar masses should thus be considered with circumspection. To minimize sources of biases, we adopted masses from \cite{dar16}, \cite{dar10}, \cite{kim16} and \cite{man12}, which all use the \cite{sie00} evolutionary model, in that order of priority. Only 11 targets in the initial sample, and only one of our observed target, has no mass estimate. As shown in Figure\,\ref{fig:masses}, the flux-limited selection results in an initial sample that is not representative of the IMF in the ONC but is heavily biased towards stars more massive than the Sun. To focus our analysis around solar-type stars, the observed sample was selected to be less dominated by intermediate-mass stars than the initial sample. The median mass in the observed sample is 0.8\,$M_\odot$, with 16 and 84 percentile at 0.4 and 1.6\,$M_\odot$, respectively. Thus, our sample is dominated by solar-type stars, albeit with the addition of a few lower and higher mass stars. About 75\% of our sample consists of K-type T\,Tauri stars and only one observed cluster members (H97\,613) has $M_\star > 2\,M_\odot$.

Finally, we used literature information to assess which of our targets possess a circumstellar disk. Specifically, we consider that a star has a disk if its SED displays significant infrared excess \citep{hil98,meg12}, if its optical spectrum reveals a strong and/or broad H$\alpha$ emission, or the infrared Ca triplet in emission \citep{hil98,sic05,fur08,dar09,man12,sze13,kim16}, or if it has an estimated accretion rate \citep{dar10}. In cases where multiple indicators of the presence of circumstellar material are available, they are in agreement with one another. The lone exception to this statement is H97\,567, which has no significant $K$ band excess \citep{hil98}, yet displays strong H$\alpha$ emission and significant Ca triplet emission \citep{hil98, dar09}. We consider that this system likely has a disk but that its near-infrared excess is too weak to be detected; no mid-infrared observations of the system are available. Of the 42 confirmed cluster members studied here, 32 are associated with a disk. Thus, our observed sample is characterized by a frequency of circumstellar disks that is consistent with the observed rate of 60--80\% in the overall ONC population \citep{hil98,lad00}.

\begin{table}
	\centering
	\caption{Observed sample. $K$ magnitudes are from the 2MASS Point Source Catalog. Spectral types and masses are from \citet{hil97} and \citet{dar16} respectively, unless otherwise noted. Additional references: $^{\it a}$ \citet{hil13}, $^{\it b}$ \citet{dar10}. The fifth and sixth column indicate whether the object posses signs of accretion and a circumstellar disk, respectively (see Section\,\ref{subsec:sample}). In a few cases, accretion indicators are ambiguous; those are indicated by a ``?'' qualifier. The last column indicates which target was previously known to be a visual binary with separation in the 0\farcs2--1\arcsec range \citep{rob13} or is a known spectroscopic binary \citep{tob09}.}
	\label{tab:sample}
	\begin{tabular}{ccccccc}
          \hline
          H97 & $K$ & Sp.T. & $M (M_\odot)$ & \multicolumn{2}{c}{Disk?} & Mult.\\
		& & & & Acc. & IR & \\
          \hline
          \multicolumn{7}{c}{Cluster members}\\
          \hline
  27    & 9.36  & K2    		& 1.43	&  N & N & \\
  29    & 9.39  & K2    		& 1.49	&  N & N? & \\
  50    & 8.60  & K1   		& 0.66	&  Y & Y & SB2 \\
 150   &  9.30  & K4--5$^{\it a}$	& 0.66	&  Y & Y &  \\
 157   &  8.06  & K2		& 1.15	&  N & N & \\
 221   &  8.01  & K3		& 1.61	&  N & N & \\
 232   &  9.22  & K1--2		& 0.61	&  Y? & Y & \\
 253   &  9.34  & K8$^{\it a}$		& 0.91	&  Y? & Y & 0\farcs27; SB2 \\
 278   &  9.32  & K2--7	& 0.98$^{\it b}$	&  Y & Y & \\
 286   &  9.14  & K5		& 1.33	&  Y & Y & \\
 337   &  9.43  & K8		& 0.64$^{\it b}$	&  Y? & Y & \\
 345   &  9.43  & M0.5		& 0.40	&  Y & Y & \\
 365   &  8.74  & K2--3	& 0.88	&  N & N? & \\
 421   &  8.62  & K5		& 1.17	&  Y & Y & \\
 423   &  8.86  & K2		& 0.39$^{\it b}$	&  Y & Y & \\
 432   &  9.25  & M3.1		& 0.33$^{\it b}$	&  Y? & Y & \\
 441   &  9.27  & M1		& 0.37	&  Y? & Y & \\
 448   &  9.14  & K7		& 0.74$^{\it b}$	&  Y & ? &  \\
 454   &  8.66  & K4		& 1.56$^{\it b}$	&  Y & Y &  \\
 460   &  8.80  & K0--3	& 1.59	&  Y & Y & \\
 478   &  8.73  & M0.4		& 0.57	&  N & N? & \\
 488   &  8.37  & K1		& 1.33$^{\it b}$	&  Y & Y & \\
 515   &  8.61  & K4--7		& \dots	&  Y & Y & \\
 529   &  9.36  & M0		& 0.56$^{\it b}$	&  Y & Y & \\
 533   &  9.42  & M0		& 0.47	&  Y & Y & \\
 534   &  9.23  & M2		& 0.39$^{\it b}$	&  Y & Y & \\
 544   &  8.19  & K4--7	& 1.97$^{\it b}$	&  N & N & \\
 550   &  8.21  & K2--3	& 1.90	&  Y & Y & 0\farcs88 \\
 567   &  7.58  & K3--4	& 1.99$^{\it b}$	&  Y & N &  \\
 596   &  7.64  & G5--K1$^{\it a}$	& 1.68	&  Y & Y &  \\
 613   &  9.04  & K2		& 2.51	&  Y? & Y & \\
 622   &  9.27  & M0--2.5	& 0.37	&  Y & Y & \\
 631   &  8.71  & K7		& 1.08	&  Y & Y & SB2 \\
 683   &  9.40  & K6		& 1.07	&  N & N & \\
 744   &  9.37  & M1$^{\it a}$		& 0.47$^{\it b}$	&  Y & Y & 1\farcs00 \\
 756   &  8.91  & M0		& 0.44	&  Y & Y & SB2 \\
 810   &  9.35  & K4		& 0.62	&  Y & Y & \\
 826   &  9.18  & K5$^{\it a}$		& 0.77$^{\it b}$	&  Y & Y & \\
 847   &  9.16  & K3		& 1.25	&  N & N & \\
3085  &  9.22  & K7		& 0.61	&  N & N & \\
3109  &  9.15  & K2--3  	& 0.61	&  Y & Y & \\
3131  &  9.45  & K5		& 1.21	&  Y & Y & \\
          \hline
          \multicolumn{7}{c}{Likely non-members}\\
          \hline
   4     &  9.43 & K4    		& 0.82	&  Y & Y & 0\farcs79; SB1 \\
  45    & 7.95  & K4  		& 1.37	&  N & N & \\
 351   &  8.79 & G4--6		& 2.43	&  N & ? & \\
 413   &  8.16 & K5		& 1.03$^{\it b}$	&  N & N? & \\
          \hline
	\end{tabular}
\end{table}


\subsection{Observations and Data Reduction}
\label{subsec:obs}

We conducted our program with the NACO instrument on VLT/UT4. The observations were conducted over 5 half-nights in January 2016 scheduled in 2 separate runs during program 096.C-0270 . All observations were made using the S13 camera (0\farcs01322/pix), with the $K_s$ filter and the 7-hole mask \citep{tut10}. Because some of our targets are faint in the visible and due to confusion from the bright nebula associated with the ONC, we used the infrared wavefront sensor mode of NACO with the N90C10 entrance dichroic to obtain optimal adaptive optics performance. 

Targets were associated in groups of 4 to 9 objects based on their magnitude and sky position in order to generate observations sequences. With this set-up, the adaptive optics parameters were set on the first target and maintained fixed from object to object, enabling rapid switching between targets. This ensures a high survey efficiency, as observing multiple science targets in rapid succession removes the need of including dedicated (single) calibrator stars. Instead, all objects found to be single stars in each sequence can serve as calibrators for the other targets. During each half-night, we executed 1 to 3 such observing sequences. In the last three half-nights, observations of some possible candidate binaries were repeated to confirm their nature, as were observations of clearly single stars that were used to serve as safe calibrators. All observing sequences are detailed in Table\,\ref{tab:log}. Integration times of 30 to 120\,s were used to ensure sufficient signal-to-noise in individual frames. Three datacubes of 4 or 6 such frames were acquired with 3--4\arcsec\ dithers between each cube to enable sky subtraction and bad pixel correction (except for H97\,4 and H97\,613, for which we only obtained 2 datacubes). 

\begin{table*}
	\centering
	\caption{Observing sequences executed during the course of this survey. Italicized targets represent non-members of the ONC.}
	\label{tab:log}
	\begin{tabular}{clc}
          \hline
	Date (UT) & H97 & DIMM seeing (\arcsec) \\
	  \hline
	01/16/2016 & {\it 45}, 157, 221, {\it 413}, 488, 544, 550, 567, 596 & 1.56--2.15 \\
		& 50, 365, 454, 478, 515, 631 & 1.20--1.63 \\
	01/17/2016 & {\it 351}, 421, 488, 756 & 0.87--1.68 \\
		& {\it 4}, 29, 337, 345, 533, 683, 3131 & 1.12--1.84 \\
	01/27/2016 & 27, 253, 278, 441, 529, 744, 810 & 0.87--1.12 \\
	01/28/2016 & 150, {\it 413}, 421, 515, 550, 567, 596 & 0.66--0.91 \\
		& 232, 286, 432, 3085, 3109 & 0.71--1.09 \\
		& 534, 622, 826, 847 & 0.71--1.09 \\
	01/29/2016 & 421, 423, 460, 478, 515, 756 & 0.67--0.77 \\
		& 432, 441, 448, 529, 622 & 0.84--0.96 \\
		& 221, 413, 544, 550, 613 & 1.08--1.26 \\
	  \hline
        \end{tabular}	
\end{table*}

Data reduction involved the usual steps of flat fielding, background subtraction and bad pixel correction. Analysis of the resulting datasets was performed in two steps. First, all images were aligned and median combined to produce ``direct'' images. In these images, the Fizeau interference pattern induced by the mask is readily evident as a combination of distinct discrete peaks, but it is still possible to identify companions outside of $\approx$0\farcs25, whose position and brightness can be determined through a cross-correlation technique. Given the number of frames per target in our observing sequence, we achieve a 5$\sigma$ contrast in the 2.5-4\,mag range. 

To identify tighter systems, however, an interferometric analysis of the data is necessary, as the signature of a companion lies in the closure phase associated with the baselines defined by the mask. To this end, we use the SAMP pipeline \citep{lac11} which decomposes the interferometric pattern in a series of discrete spatial frequencies (each defined by a unique pair of holes) and computes the closure phases for each distinct triangle of holes from the corresponding bispectrum. Those closure phases, which should be null for a point source, are calibrated by subtracting the average closure phase observed for all single source in each observing sequence. We then fit a single star model and a binary system model to all data on a given target. The binary star model is selected only in cases where the $\chi^2$ of the single star model is unacceptable. Otherwise, a map of the 5$\sigma$ detection limit is produced for each target. This results in a roughly separation-independent detection limit between 0\farcs04 and 0\farcs15, where the outer search radius is set by the diffraction limit corresponding to the shortest spacing between the mask holes. The 5$\sigma$ sensitivity of our aperture masking survey ranges from 2.5 to 4\,mag, similar to the sensitivity achieved by direct imaging at larger separation, as discussed above (see Figure\,\ref{fig:comps}). At the closest separations, the detection limit degrades gradually down to $\approx$0\farcs02, inside of which sensitivity to companion vanishes in aperture masking. 


\section{Results}
\label{sec:res}


\subsection{Detected companions}
\label{subsec:comps}

The observed properties of all companions are listed in Table\,\ref{tab:comps}. Inspection of the direct images revealed only 2 companions which had already been discovered in HST images of the ONC with relative astrometry and photometry consistent with our results \citep{rob13}. On the other hand, we did not detect the HST-detected companions to H97\,550 and H97\,744. The former companion is $\approx 6.5$\,mag fainter than its primary in the red portion of the visible and thus well below our detection limit in the near-infrared. The latter is about 2.5\,mag fainter than the primary in the near-infrared, but our detection limit for that source is $\Delta K_s \approx 2.5$\,mag, so that the non-detection is still consistent with previous knowledge of the system.

The primary driver of this study is the search for closer companions. Our closure phase analysis resulted in the discovery of 13 companions, with separations ranging from 0\farcs023 to 0\farcs151 and contrast ratios as high as 4.3\,mag. The companions to H97\,50 and H97\,567 are located at the edge of the range of separations probed by aperture masking and their properties are affected by large uncertainties. Nonetheless, we consider them as real companions as they consistently appear when we use different subsets of calibrators to analyze the datasets for these sources. The companion to H97\,567 was also confirmed through its detection in two distinct observations. We also note that this latter binary, with a projected separation of about 9\,au, is the only disk-bearing system with no near-infrared excess, suggesting that the disk could be circumbinary in nature, with only modest amount of circumstellar material, while still allowing accretion streamers on the central sources.

Confusion between physically bound companions and chance projection of unrelated stars (another cluster member or a fore/background star) has always been a serious concern in multiplicity studies of the ONC. Using the star count computed for the core of the ONC by \cite{koe06} and integrating down to $K \approx 12.5$, or 3\,mag deeper than our fainter primaries, we conclude that there is a 0.3\% probability of chance alignment with an unrelated star within the 0\farcs15 outer radius of our aperture masking search space for any one target. Over the whole sample, this results in a 10\% probability that there is one such pair among the candidate companions we have identified. As could be expected, given the very small angular scale over which we are searching for companions, this is an unlikely event and we therefore assume from now on that all candidate companions are physically associated to their primaries.

\begin{table}
	\centering
	\caption{Close companions detected in this survey. Notes: $^{\it a}$ These companions are at the edge of the range accessible through aperture masking and thus their measurements are associated with large uncertainties; $^{\it b}$ These companions were already identified in previous HST optical images \citep{rob13}.}
	\label{tab:comps}
	\begin{tabular}{cccc}
          \hline
	H97 & $\rho$ (mas) & PA (\degr) & $\Delta K$ (mag)\\
	  \hline
	\multicolumn{4}{c}{Closure phase analysis} \\
	  \hline
  50$^{\it a}$ & 151.3$\pm$10.0 & 326.3$\pm$  5.1 & 2.44$\pm$1.86 \\	
232 &  57.5$\pm$3.3 & 306.5$\pm$2.6 & 1.36$\pm$0.12 \\	
253 &  86.4$\pm$6.0 &  93.1$\pm$7.2 & 2.12$\pm$0.08 \\	
286 &  67.7$\pm$13.4 & 226.1$\pm$11.3 & 1.71$\pm$0.34 \\	
345 &  56.2$\pm$3.4 & 358.0$\pm$4.2 & 1.12$\pm$0.01 \\	
432 &  54.9$\pm$3.7 & 323.5$\pm$3.0 & 1.22$\pm$0.09 \\	
441 &  48.4$\pm$4.6 & 216.6$\pm$7.0 & 2.32$\pm$0.19 \\	
488 & 130.7$\pm$5.4 & 262.4$\pm$1.1 & 2.59$\pm$0.34 \\	
550 &  30.2$\pm$1.6 &   3.9$\pm$17.2 & 0.97$\pm$0.03 \\	
567$^{\it a}$ &  22.7$\pm$14.5 & 114.0$\pm$20.3 & 1.27$\pm$3.67 \\	
596 &  74.1$\pm$6.4 &  18.6$\pm$13.4 & 4.28$\pm$0.53 \\	
683 &  88.9$\pm$7.6 &  64.5$\pm$3.5 & 3.19$\pm$0.32 \\	
3131 &  96.2$\pm$8.7 &   6.5$\pm$4.8 & 3.20$\pm$0.67 \\	
	  \hline
	\multicolumn{4}{c}{Inspection of direct images} \\
	  \hline
4$^{\it b} $ & 805$\pm$20 &  212$\pm$2 &   2.04$\pm$0.11 \\
253$^{\it b}$ &  283$\pm$13 &  359$\pm$2 &  0.63$\pm$0.08 \\
	  \hline
        \end{tabular}	
\end{table}

\begin{figure}
    \includegraphics[width=\columnwidth]{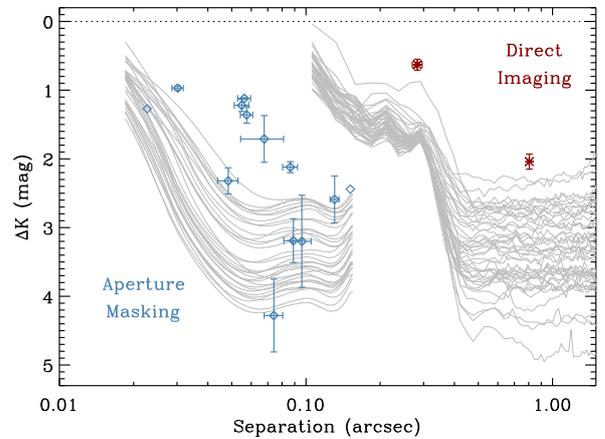}
    \caption{Detected companions and individual 5$\sigma$ detection limits for targets in our survey; targets with a detected companion within the range of separation of each method are excluded as their detection limit are significantly affected by the presence of a second point source. Blue diamonds and red asterisks represent companions detected by closure phase analysis and cross-correlation of direct images, respectively. Black diamonds mark two companions detected in the closure phase analysis but whose flux ratio is poorly estimated (see Section\,\ref{subsec:comps}).}
    \label{fig:comps}
\end{figure}

All companions detected in this survey are shown in Figure\,\ref{fig:comps} along with our individual 5$\sigma$ detection limits. Interestingly, we detected no companion with $\Delta K_s \lesssim 1$\,mag. While this could indicate a dearth of nearly equal-mass binaries, it is important to note that the presence of thermal emission from circumstellar disks (present in the majority of the systems targeted here) can significantly alter the near-infrared brightness of young stars. For similar reasons, we refrain from converting the $K_s$ flux ratio into a mass ratio as uncertainties on the primary masses and contamination from disk emission are large effects that cannot be satisfyingly handled with single-wavelength observations. We do note, however, that several companions have an apparent magnitude that is $K > 11.3$, which is the predicted brightness of an unextincted, 1 Myr-old 0.08\,$M_\odot$ object at the distance of the ONC based on the evolutionary models of \cite{all12}. In particular, the companions to H97\,683 and H97\,3131 are more than 1\,mag fainter than this limit, making them candidate brown dwarf companions. The fact that high line-of-sight extinctions are common in the ONC raises caution, however. Extinctions as high as $A_V\approx10$\,mag are found in the ONC \citep{dar16}. We defer further discussion of the mass ratios of the detected systems and of the nature of these apparently extremely faint companions until further photometric and spectroscopic characterization has been obtained.

Finally, we note the presence of two apparent high-order multiple systems in our sample. We found a close companion to the 0\farcs88 binary H97\,550; the ratio of projected separations in the systems is 29, ensuring that it is most likely dynamically stable in the long term. On the other hand, the situation for the H97\,253 system is complicated: not only was it already known as both a spectroscopic and visual binary (see Table\,\ref{tab:sample}), but our survey discovered a new 0\farcs086 companion. This companion is highly unlikely to be the same as the spectroscopic companion since the latter is characterized by a relative radial velocity of at least 10\,km\,s$^{\mathrm -1}$, i.e., with a semi-major axis that is likely smaller than $\approx$15\,au, or $\approx$0\farcs025. H97\,253 could therefore be a quadruple system. To be stable on the long-term, hierarchical systems must have a ratio of semi-major axes that exceeds $\approx$3, although the exact threshold is dependent on the eccentricity, mass ratio and relative inclination of the sub-pairs \citep{mar99}. With a single epoch of observation and without any knowledge of the extent of projection effects, it is currently impossible to assess the long-term stability of this system, however.


\subsection{Multiplicity properties}
\label{subsec:mult}

We focus our multiplicity survey on the 10--60\,au (0\farcs026--0\farcs155) projected separation range, which has not been probed in previous surveys of the ONC and where we have near-uniform sensitivity. In this range, we identified 12 companions to 42 targets, for a raw CSF of 28.6$^{+7.8}_{-5.9}\,$\% (68 percentile uncertainties are computed using binomial statistics). A classical issue inherent to multiplicity surveys based on flux-limited samples is the Branch bias that leads to an over-representation of faint binaries. Indeed, the brightness of a binary or high-order multiple system can be sufficient for survey inclusion even though no single star in the system exceeds the threshold. From the system $K$ magnitude and our measured flux ratios, we determined that four systems (H97\,253, H97\,345, H97\,432 and H97\,3131) were included as a result of this bias. Discounting these objects, our surveys revealed 8 companions to 38 targets, for a CSF of 21.1$^{+8.0}_{-5.1}\,$\%. We note that because our sample definition also included a maximum brightness, it is possible that some systems with a primary in our $K_s$ range but with a companion ended up being excluded from the survey in an ``anti-Branch bias.'' Given the small numbers of ONC targets lying a few tens of a magnitude brighter than our $K=7.5$ upper threshold, though, few systems are likely to be affected in this way. The true CSF in the ONC is therefore likely to be only slightly higher than this estimate.

Figure\,\ref{fig:comps} shows that most of our companions lie above the 5$\sigma$ detection limit for all single stars, and all but one are brighter than the median detection limit. This suggests that the completeness of our survey to companions is high, at least down to $\Delta K \approx 3$\,mag. It is possible that a handful of companions with $\Delta K \gtrsim 2$\,mag and projected separations smaller than 0\farcs04 could have been missed, as well as faint ($\Delta K \gtrsim 4$\,mag) companions over most separations. However, evaluating the amplitude of this effect requires making assumptions about the distributions of flux ratio and separation as well as their covariance. We feel that the number of companions discovered in our survey is insufficient to enable accurate estimates and chose not to apply a completeness correction. In turn, this means that the companion frequency found in this survey is a conservative lower limit to the actual one.

We fail to identify any significant dependency of stellar multiplicity within our sample. The binary systems possess similar distributions of $K$ magnitude, spectral type and estimated primary mass as the observed sample, and their spatial distribution in the cluster is also indistinguishable from that of single stars (see, Figure\,\ref{fig:distance}, \ref{fig:kmag} and \ref{fig:masses}). We conclude that our estimated CSF applies to the ONC as a whole, at least out to 2\,pc from $\theta^1$\,Ori\,C.

At first glance, there appears to be an excess of companions among disk-bearing targets (22 and 28\% with and without correction for the Branch bias, respectively) over diskless targets (10\%). However, the small number of targets in the latter category - there is only 1 binary in that subsample - leads to large uncertainties and the difference is not statistically significant. Nonetheless, this result is surprising since visual companions with separation smaller than 40--50\,au have previously been found to be predominantly associated with disk-free T\,Tauri stars in other SFRs \citep{cie09, kra12,che15}. These past studies, however, considered several nearby SFRs but did not include Orion due to the inability to identify such close companions at this larger distance. This could indicate that disk formation and survival in close binaries proceeds differently in a dense cluster like the ONC compared to other SFRs, or that the disk survival time in close binaries is similar to, or slightly larger than, the age of the ONC cluster. In the latter scenario, disks would need to dissipate quickly beyond that phase in order to match the results derived from other star-forming regions.


\subsection{Comparison to other surveys}
\label{subsec:compar}

In order to place our results in context, we must now compare the CSF to that observed among field stars and other young stellar populations. Among field solar-type and low-mass stars, the CSF in the 10--60\,au range are 11.7$\pm$1.6\% and 6.5$\pm$1.6\%, respectively \citep{rag10,war15}. The CSF we found in the ONC is much higher, roughly twice as high as the field solar-type stars, the more appropriate comparison sample given the make-up of our observed sample. However, owing to small number statistics in our survey, the statistical significance of the difference is not definitive: the excesses over solar-type and low-mass stars are significant at the 91.8 and 99.3\% confidence levels (1.7$\sigma$ and 2.7$\sigma$), respectively. Nonetheless, this is the first tantalizing evidence for an excess of multiple systems in the ONC over field stars.

The observed CSF in the Taurus, Ophiuchus, Upper Scorpius SFRs and the $\beta$\,Pic Moving Group (BPMG) over the same separation range are approximately 22, 16, 16.5 and 19\%, respectively \citep{kra08,kra11,che15,ell16}. These are approximate rates, as complex object-dependent completeness corrections have been applied in each of these surveys, but the amplitude of these corrections in our separation range is modest and consistently smaller than the statistical uncertainties, which are typically $\pm$3--5\%. The CSF we have measured in the ONC is consistent with those observed in other young stellar populations and, if anything, closer to that observed in Taurus, which has the highest CSF in nearby SFRs. 

\begin{figure*}
    \includegraphics[width=0.7\textwidth]{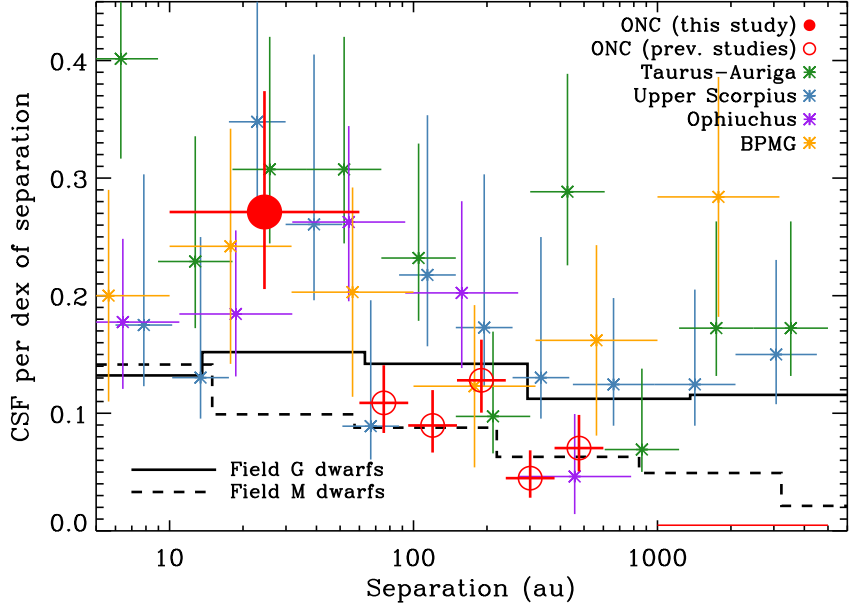}
    \caption{Separation distribution for multiple systems observed among field stars and nearby SFRs. In each bin, the observed CSF is normalized by decade of projected separation to enable direct comparisons between surveys probing different bin sizes. The distribution in the ONC is shown as red circles \citep[this survey as the filled circle and][]{rei07} and an upper limit at the widest separations \citep{sca99}, whereas the corresponding distributions for the low-mass and solar-type stars in the Taurus-Auriga, Upper Scorpius, Ophiuchus SFRs and in the BPMG are shown as asterisks \citep{kra08,kra09,kra11,che15,ell16}. The distributions for G and M dwarfs (continuous histograms) are taken from \citet{rag10} and \citet{war15}, respectively.}
    \label{fig:sep_dist}
\end{figure*}

Figure\,\ref{fig:sep_dist} illustrates the separation distribution observed in the ONC, other young stellar populations and among field stars. For the ONC, we adopted the results of \cite{rei07} for separations larger than 60\,au as it is the largest survey to date. In most SFRs, the observed distribution of separations is broad, consistent with the log-normal distribution observed among field stars \citep{rag10,war15}. Indeed, such a parametrization has been successfully used in SFRs \citep[e.g.,][]{kra12,che15}. In the ONC, on the other hand, we find a sharp decline in the CSF outside of $\approx$60\,au, although we do not have sufficient statistical strength to tightly constrain this threshold separation. While Taurus and the ONC have undistinguishable CSFs in the 10--60\,au range, Taurus has 2.5 times more companions in the 60--150\,au range. Furthermore, the sharp decline around 60\,au identified in this study contrasts with the rather shallow separation distribution between 60 and 600\,au, suggesting that the shape of the separation distribution is intrinsically different in the ONC compared to other SFRs and to the field population. 

Finally, since the ONC is a plausible precursor to Pleiades-like clusters, it is meaningful to compare the companion fraction we observe in the ONC to that of nearby open clusters. \citet{bou97, bou01} and \citet{pat98,pat02} probed the visual multiplicity of solar-type stars in the Pleiades, Hyades, Praesepe and $\alpha$\,Per clusters. While these studies probed separations comparable to those we consider here, their sensitivity to low-mass companions was limited to companions with mass ratios $\gtrsim 0.3$--0.4 in this range as a consequence of the older ages of these clusters. These studies applied completeness corrections to alleviate this problem, but this introduces significant uncertainties as the correction factors are large \citep[e.g., a factor of 4 in the 14-50\,au range in the Pleiades;][]{bou97}. \cite{pat02} produced a global analysis of all four open clusters, concluding that the frequency of visual companions (26--581\,au) in these environments is similar to that of field stars. However, their analysis also showed that the distribution of projected separations is skewed towards tighter separations than in the field, with a peak at $\approx4$\,au, i.e. a factor $\approx10$ tighter than among field stars. This suggests that open clusters are characterized by a relative deficit (alternatively, excess) of companions at hundreds of au (alternatively, tens of au and tighter). The statistical and systematic uncertainties in the derived separation distribution are too large to allow for a definitive comparison with the results of this survey, however.


\section{Discussion}
\label{sec:disc}


\subsection{Is the close multiplicity excess in the ONC real?}

Taken at face value, our survey has revealed that solar-type members of the ONC host more companions in the 10--60\,au range than their field counterparts, the first time such a multiplicity excess is identified in that region. Indeed, the CSF for tight companions in the ONC population is consistent with that observed in other SFRs, contrary to what was found at larger separations over the last two decades. If confirmed, this has profound implications for our understanding of the process through which multiple system form and to the star formation process at large. Before discussing these implications, it is necessary to evaluate the possibility that the main conclusion of this survey is  skewed by uncorrected biases. The most obvious bias associated with multiplicity survey is the Branch bias, which we have corrected for. Hence more subtle biases must be considered.

First of all, we evaluate whether our observed sample is biased relative to the initial sample from which it was drawn. The spatial distributions of the two samples conform well to one another (see Figures\,\ref{fig:map} and \ref{fig:distance}), with the caveat that our survey under-represents the NE region of the ONC relative to the S and E outskirts of the cluster. Baring a major dynamical anisotropy in the cluster's dynamics, we consider it unlikely that this can significantly affect our analysis. While our sample extends out to 2\,pc from the cluster's center, half of our targets are located within 0.3\,pc of the Trapezium (Fig.\,\ref{fig:map}). In other words, our survey primarily focuses on the core of the cluster and we have to consider the possibility that this is a sub-population with an elevated multiplicity frequency. For instance, mass segregation has been identified in the ONC for high-mass stars \citep{hil98b} and for brown dwarfs \citep{and11}. While the origin of the former is still debated, the latter is most likely a consequence of the dynamical evolution of the cluster, which expels preferentially its lowest-mass members. It is conceivable that this same mechanism preferentially ejects single stars \citep{del97}, thus leading to a remaining population that has an elevated CSF compared to its initial value. The fact that multiple systems are not more centrally condensed than single stars within our survey suggests that this is not a significant effect. Indeed, we computed the Minimum Spanning Tree \citep{kru56} of both the singles and binaries subsamples, and their mean branch lengths are indistinguishable at the 1$\sigma$ level. Furthermore, if a widely dispersed of primarily single stars were now present in the outer regions of the ONC, it would imply that all CSF estimates for that region have so far been overestimated, not just for a specific separation range. The multiplicity survey of \citet{rei07} covered a very similar area to ours, for instance. Thus, if this were the case, we would conclude that the ONC population has a much lower CSF than the field outside of 60\,au, thereby introducing a new mismatch between the ONC and field populations. 

Second, the observed subsample is not significantly biased in terms of brightness compared to the initial sample (Figure\,\ref{fig:kmag}), even after accounting for the 4 systems that were included because of the Branch bias. Besides, in all likelihood some unobserved members of the initial sample also are unresolved binaries that would not meet the minimum brightness criterion based on the brightness of their primary alone. 

One possible bias associated with our survey is related to the presence of circumstellar disks in the majority of the systems targeted in this survey. Based on observations of other star-forming regions, this could potentially introduce a bias towards a lower binary companion (see Section\,\ref{subsec:mult}. Possible issues in assessing the presence of a disk (crowding, contamination from the surrounding nebula) as well as the unknown survival time of disks in close binaries prevent us from evaluating the amplitude of this bias, but we conclude that it can only further strengthen the significance of the multiplicity excess in the ONC compared to field stars.

Finally, while the observed sample has a deficit of stars with $M_\star \gtrsim 1.25\,M_\odot$ relative to the initial sample (Figure\,\ref{fig:masses}), this is by design so that we can realistically compare our results to surveys of solar-type stars in other environments. Indeed, surveys in nearby SFRs typically include stars with a range of masses that is broader and extends to lower mass than our survey in the ONC, and thus these should in principle be best compared to a weighted average of the field solar-type and low-mass stars. However, none of the surveys listed above found strong mass dependencies of the CSF for visual binaries, nor do wee see a significant one in this survey (see Figure\,\ref{fig:masses}). Thus, the comparison between SFRs remains valid. Either way, the CSF observed in the ONC for 10--60\,au is well above that observed in the field for both solar-type and lower-mass stars. One conceivable way to ascribe the multiplicity excess we find to an underlying stellar mass bias would be if stellar masses in the ONC have been consistently underestimated by a significant amount, so that a significant fraction of our sample consists of intermediate-mass stars. The latter are thought to host a higher frequency of close visual companions \cite[albeit with large uncertainties in the separation range under consideration here;][]{riz13,der14}. This seems difficult to reconcile with the spectral type of the targets in our sample, however, as 2\,$M_\odot$ stars are expected to be in the mid-G spectral type range according to most evolutionary models \citep[e.g.,][]{man12}. On the basis of the available data, we thus exclude that our sample is strongly affected by intermediate-mass stars.

In summary, no significant bias appears to be skewing the conclusions of our survey, and thus we confirm that 1) solar-type members of the ONC host an elevated CSF -- by a factor of almost 2 -- in the 10--60\,au range compared to field stars, and 2) that the CSF observed in the ONC is fully consistent with that observed in other SFRs. We now turn our attention to the implications of these findings.


\subsection{Long-term stability of ONC close binaries}

Binaries with semi-major axes of just a few tens of au are stable over billions of years once they are released in the Galactic field \citep{wei87}. Thus, if the excess of close binaries in the ONC is a temporary feature, whereby some of these systems will either break apart or significantly change their orbital period, it must be as a consequence of processes internal to the ONC and/or to the multiple system itself. We address both of these possibilities here. To reconcile the observed CSF in the ONC with that of the field, roughly half of the 10--60\,au companions range must be removed from that range.

There are numerous indicators that the ONC is a dynamically rich environment. The lack of very wide binaries \citep{sca99} and the apparent deficit of binaries wider than 200\,au in the inner pc of the cluster \citep{rei07} are likely indicative of dynamically violent interactions affecting multiple systems in the cluster. It is therefore worth exploring whether the close binaries identified in this survey can survive the long-term evolution of the cluster until its dissolution in the field. The dynamical state of the ONC is not firmly established; it may be expanding -- in the initial phases of dissolution -- or close to virial equilibrium \citep{all09,tob09,dar17,kro18}. Either way, the cluster was (much) denser in the past and, as a consequence, most disruptive interactions occurred earlier in its evolution \citep[e.g.,][]{kro99}. As a rule of thumb, a binary system will get destroyed by a passing third body if the relative velocity of the encounter is equal to the orbital velocity  of the binary \citep{hil90}. Assuming random directions for the traveling directions of systems, the encounter velocity can be approximated as twice the velocity dispersion of the population. Given the current velocity dispersion in the cluster \citep[$\approx$2\,km\,s$^{-1}$;][]{dar17}, this implies that systems with orbital velocities of $\gtrsim$4\,km\,s$^{-1}$ can survive contemporary and future interactions in the cluster. Assuming a mean system mass of 1.5\,$M_\odot$ and circular orbits, this orbital velocity corresponds to a semi-major axis of $\approx$80\,au. Therefore, we conclude that the close binary systems studied here are stable against the future evolution of the ONC. 

An alternative mechanism to dynamically alter the close binaries we have identified is related to the evolution of compact three-body systems. If such systems are not hierarchical, i.e., when the ratio of the outer and inner semi-major axes is $\lesssim3$, mutual interactions typically lead to a tightening of the inner pair and a corresponding expansion of the outer orbit, sometimes up to the point of instability and ejection. The timescale for this evolution depends on the initial separations, and could be on the order of a few Myr for systems similar to those we are probing in the ONC \citep[e.g.,][]{rei12}. Thus, it is possible that some of the binaries we have identified will evolve significantly before the cluster is fully dissolved, crucially displacing the companions to outside the 10--60\,au range. We have only identified two high-order multiple systems, but it is plausible that some high-order systems are still unaccounted for. It is unlikely that such missing companions would be located at larger separation, as direct imaging can readily detect any stellar companion to a solar-type ONC member. Therefore, for the ``unfolding triple system'' scenario to account for the apparent excess of 10--60\,au companions, the missing companions must be closer in, at separations of a few au. However, the distribution of separation declines at separations of $\lesssim$10\,au for both the field population and in SFRs \citep[e.g.,][]{rag10,ell15}, and the CSF observed among solar-type field stars in the 1--10\,au range is only about 12\%. It is therefore unlikely that several of the binary systems identified here also possess a closer in third component that could significant affect the orbit of the detected companion.

In summary, the present and future dynamical states of the ONC, as well as the likely proportion of high-order multiple in our sample, appear insufficient to effectively remove many 10--60\,au companion. Thus, the elevated CSF we have found in the ONC will remain mostly unchanged as the cluster is dissolved into the Galactic field.


\subsection{Implications}

Our survey has revealed that solar-type members of the ONC host twice as many companions in the 10--60\,au range as their field counterparts at a high confidence level, the first time such a multiplicity excess is identified in that SFR. Indeed, the CSF for tight companions in the ONC population is consistent with that observed in other SFRs, contrary to what has already been documented at larger separations. Furthermore, the distribution of orbital separation in the ONC is characterized by a sharp drop-off outside of 60\,au that is unlike what is seen in other populations, either in SFRs or in the field. We now discuss how these findings affect our understanding of star formation at large.

While the results of this survey cannot definitively solve the ``nature vs nurture'' debate regarding multiplicity, the fact that all SFRs that have been probed to date shares a similar CSF over the 10--60\,au range is more naturally consistent with the hypothesis of a universal set of initial multiplicity properties. Indeed, calculations by \cite{kro01} and \cite{par12} tailored to reproduce the occurrence of wider binaries in the ONC and based on Taurus-like initial conditions predict a marked excess in the ONC over field stars at separation $\lesssim$100\,au, in good agreement with our findings. Fundamentally, binaries tighter than 60\,au are too hard to be significantly affected by the past evolution of the cluster. Furthermore, the predicted sharp decline with increasing separation out to 1000\,au and the absence of even wider systems is fully consistent with all observations of the ONC. While it remains speculative to trace back the population of wider systems in the ONC since it depends on the dynamical history of the cluster, our survey was designed to probe pristine multiple systems, i.e., systems that have not been affected by this prior evolution. Thus, the match in CSF between the ONC and other SFRs indicates that, at least for the 10--60\,au separation range, star formation proceeds to a near-universal CSF irrespective of the region. 

In turn, this implies that the global properties of a giant molecular cloud play a negligible role in the formation of multiple systems, since the relatively quiescent environment of the Taurus SFR, for instance, is dramatically different from the ONC. Instead, our results suggest that the formation multiple system depends primarily on local conditions, and that these conditions must be sufficiently similar in all SFRs. For instance, this could happen if some self-regulatory process leads to prestellar cores that are comparable in all environments, leading them to fragment in a similar fashion. This is qualitatively consistent with effect of cloud turbulence, whose amplitude and power spectrum only mildly affect the resulting multiplicity properties \citep{del04,bat09}. Conversely, the influence of magnetic field and radiative feedback appears more significant, albeit this is still an ongoing debate \citep[e.g.,][]{hen08,pri08,off09,bat12,lom15}. The question of whether and how cloud formation and collapse can self-regulate, thus leading to a universal set of multiplicity properties remains open, and is beyond the scope of our study. We note however, that observed properties of prestellar cores in isolated situations (e.g., in the Taurus SFR) differ in size, density and level of turbulence from those found in more clustered environments \citep{war07}, possibly indicating that the self-regulation process is enacted after the formation of the prestellar cores.

While our findings support a near-universal set of initial multiplicity properties, this renews the question of the origin of field stars. Previous observations of multiple systems in the ONC, on scales of a few hundred au, were consistent with the field and, thus, the idea that the field is primarily populated by stars that have formed in similar, or slightly looser, clusters \citep[e.g.,][]{kro95,pat02}. Our results now exclude this scenario given the observed excess of companions in the 10--60\,au range. Indeed, since a wide range of SFRs share the same CSF in this range, if the galactic field was primarily populated from SFRs like the ONC or less dense ones, there would be twice as many tight companions in the field population as is actually observed. One possible solution to this problem is to assert that most field stars form in yet denser clusters than the ONC, which can effectively destroy even the close visual binaries we probed in this study. This is problematic at two levels, however. First of all, while studies based on cluster counts favor the idea that clusters of a broad range of sizes contribute to star formation in the Solar neighborhood, they are dominated by clusters that are less dense and rich than the ONC, not denser and richer \citep[e.g.,][]{ada01,bre10,war18}. The steep power law slope of the mass distribution of stellar clusters \citep[][and references therein]{ada10} also refutes the idea that most field stars arise from very rich clusters.

Second, such dense clusters have the ability to destroy essentially all binaries wider than 100--200\,au, which would introduce a different but equally problematic mismatch with the field population. This issue is actually a profound one. In short, a given initial cluster density results in a final orbital period distribution that is a truncated version of the initial one, with a sharp decline around the ``destruction limit'' (corresponding to about 60\,au in the ONC). While this is consistent with observations of the ONC and of much lower density environments such as Taurus, the field population is characterized by a broad distribution of orbital periods that cannot be reproduced by a linear combination of cluster densities under the assumption of universal initial multiplicity properties. Indeed, the necessity of a large fraction of stars formed in relatively loose environments to account for the rich population of wide binaries in the field would in turn result in a much higher CSF at shorter separation that is inconsistent with the field population.

In summary, the field population of solar-type multiple systems cannot be accounted for by the dynamical evolution of a universal initial population, even if one considers a broad diversity of star-forming environments that spans the range from regions like the Taurus association and the core of the ONC (Ophiuchus and Upper Scorpius are intermediate in richness and density between these two extremes). Multiplicity surveys for similarly close binaries in other Orion sub-regions, such as the outer ONC, the low-density L1641 cloud and the NGC\,2024, 2068 and 2071 clusters, would be most valuable to test whether the universality holds throughout Orion. Instead, it is possible that field stars form majoritarily in environments that are not well represented by the SFRs located within 500\,pc of the Sun and, crucially, that these environments would give birth to a population of multiple systems that is significantly different. In other words, we are led to the paradoxical conclusion that, while nearby SFRs are consistent with a universal output of multiple systems, this does not apply to other environments which must nonetheless account for a majority of field stars. The notion that nearby SFRs are not representative of star formation on Galactic scales is uncomfortable, given that we rely on these regions to inform our current understanding of star formation. Since solar-type field stars are several Gyr-old on average, it could be that the output of star formation in the past led to a universal-but-different set of multiplicity properties, possibly as a consequence of the lower metallicity in the clouds that produced these older stars. Since core fragmentation is a consequence of the so-called opacity limit, which marks the phase when a collapsing core becomes optically thick and can no longer effectively cool \citep{lar69,mas00}, one expects the metallicity of the initial cloud to be an important physical factor as is sets the amounf of dust it contains. Qualitatively, lower metallicity cores should be capable of collapsing further before fragmenting -- if at all -- thereby producing less binaries on the scales of tens of au, the typical fragmentation scale in present-day clouds. A tentative dependency on total multiplicity with metallicity has been suggested among field stars \citep[e.g.,][]{rag10}, although there is still ambiguity in the interpretation due to complex biases \citep{duc13}. Whether such a metallicity dependence on core fragmentation is at the root of the difference in multiplicity properties between field stars and young stellar populations remains an open question for now.


\section{Conclusions}
\label{sec:concl}

We have conducted a near-infrared survey for close visual binaries among 0.3--2\,$M_\odot$ members of the ONC using the aperture masking technique on the 8m VLT telescope. This method allows us to probe for the first time the frequency of companions at separations $\leq60$\,au in this cluster. Out of 42 targets, we have identified 13 new companions. Previous surveys in the ONC, which focused on wider projected separations, have consistently found that the multiplicity in the cluster is consistent with that of the galactic field population and roughly half as high as observed in other nearby SFRs. In marked contrast, we find a CSF in the 10--60\,au range of 21$^{+8}_{-5}$\%, which is consistent with other SFRs and roughly double that observed among field-stars after correcting for the Branch bias. Compared to field stars, this excess is significant at the 92--99\% level. We find no clear dependency of multiplicity as a function of stellar properties or location in the ONC. Surprisingly, since our sample is dominated by disk-bearing targets, our results suggest that the disruptive effect of close binaries on disk survival are not as marked in the ONC as in other SFRs, or that these effects have not yet reached their full scale. The match in CSF between the ONC and other SFRs, together with the sharp decline towards larger separations is consistent with the hypothesis of a universal set of multiplicity properties in all SFRs coupled with intra-cluster dynamical evolution. This would indicate that the fragmentation process that gives rise to visual binaries is largely independent on the global properties of the parent molecular cloud and that the local physical properties are sufficiently self-regulated so as to proceed in similar fashion in dense clusters and quiescent associations. In addition, the results of our survey renews the question of the origin of field stars, as the close binaries we identified in the ONC will not be destroyed during the remainder of the cluster dissolution. Thus, if most stars in the field arise from regions similar to, or less dense than, the ONC, they would host a higher frequency of close visual binaries. This may indicate that nearby SFRs are not representative of the conditions that reigned when the majority of field stars formed, several Gyr ago. 


\section*{Acknowledgements}

The authors are grateful for helpful conversations about the implications of our results with Matthew Bate, Patrick Hennebelle, Isabelle Joncour, Pavel Kroupa, Charles Lada, Michael Marks and Hans Zinnecker, to the ESO staff for their help in conducting the observations presented in this study, and to a rapid and encouraging report from an anonymous referee. This work was supported in part by the Agence Nationale pour la Recherche under grant 2010-JCJC-0501-1 ``DESC'' (Dynamical Evolution of Stellar Clusters) and by ESO. EM acknowledges financial support from the "StarFormMapper" project funded by the European Union's Horizon 2020 Research and Innovation Action (RIA) program under grant agreement number 687528. SL acknowledges support from ERC starting grant No. 639248. This research has made use of the SIMBAD database and of the VizieR catalogue access tool, operated at CDS, Strasbourg, France.





\begin{thebibliography}{99}
\bibitem[\protect\citeauthoryear{Adams \& Myers}{2001}]{ada01} Adams, F.~C., \& Myers, P.~C.\ 2001, ApJ, 553, 744 
\bibitem[\protect\citeauthoryear{Adams}{2010}]{ada10} Adams, F.~C.\ 2010, ARA\&A, 48, 47 
\bibitem[\protect\citeauthoryear{Allard et al.}{2012}]{all12} Allard F., Homeier D., Freytag B., Sharp C~ M. 2012, in {\it EAS Publ. Ser.} 57, Reyl\'e C., Charbonnel C., Schultheis M. eds., 3
\bibitem[\protect\citeauthoryear{Allison et al.}{2009}]{all09} Allison, R.~J., Goodwin, S.~P., Parker, R.~J., et al.\ 2009, ApJL, 700, L99 
\bibitem[\protect\citeauthoryear{Andersen et al.}{2011}]{and11} Andersen, M., Meyer, M.~R., Robberto, M., Bergeron, L.~E., \& Reid, N.\ 2011, \aap, 534, A10 
\bibitem[\protect\citeauthoryear{Bate}{2009}]{bat09} Bate, M.~R.\ 2009, MNRAS, 397, 232 
\bibitem[\protect\citeauthoryear{Bate}{2012}]{bat12} Bate M.~R., 2012, MNRAS, 419, 3115
\bibitem[\protect\citeauthoryear{Bouvier et al.}{1997}]{bou97} Bouvier J., Rigaut F., Nadeau D. 1997, A\&A, 323, 139
\bibitem[\protect\citeauthoryear{Bouvier et al.}{2001}]{bou01} Bouvier J., Duchene G., Mermilliod J.~C., Simon T. 2001, A\&A, 375, 989
\bibitem[\protect\citeauthoryear{Bouy et al.}{2014}]{bou14} Bouy H., Alves J., Bertin E., Sarro L.~M., Barrado D. 2014, A\&A, 564, A29
\bibitem[\protect\citeauthoryear{Bressert et al.}{2010}]{bre10} Bressert E. et al. 2010, MNRASL, 409, L54
\bibitem[\protect\citeauthoryear{Cheetham et al.}{2015}]{che15} Cheetham A.~C., Kraus A.~L., Ireland M.~J., Vieza L., Rizzuto A., Tuthill P.~G. 2015, ApJ, 813, 83
\bibitem[\protect\citeauthoryear{Cieza et al.}{2009}]{cie09} Cieza L.~A. et al. 2009, ApJL, 696, L84
\bibitem[\protect\citeauthoryear{Da Rio et al.}{2009}]{dar09} Da Rio N. et al. 2009, ApJS, 183, 261
\bibitem[\protect\citeauthoryear{Da Rio et al.}{2010}]{dar10} Da Rio N. et al. 2010, ApJ, 722, 1092
\bibitem[\protect\citeauthoryear{Da Rio et al.}{2016}]{dar16} Da Rio N. et al. 2016, ApJ, 818, 59
\bibitem[\protect\citeauthoryear{Da Rio et al.}{2017}]{dar17} Da Rio N. et al. 2017, ApJ, 845, 105
\bibitem[\protect\citeauthoryear{de La Fuente Marcos}{1997}]{del97} de La Fuente Marcos, R.\ 1997, A\&A, 322, 764 
\bibitem[\protect\citeauthoryear{Delgado-Donate et al.}{2004}]{del04} Delgado-Donate E.~J., Clarke C.~J., Bate M.~R. 2004, MNRAS, 347, 759
\bibitem[\protect\citeauthoryear{Duch\^ene}{1999}]{duc99} Duch\^ene G. 1999, A\&A, 341, 547
\bibitem[\protect\citeauthoryear{Duch\^ene et al.}{1999}]{duc99b} Duch\^ene G., Bouvier J., Simon T. 1999, A\&A, 343, 831
\bibitem[\protect\citeauthoryear{Duch\^ene \& Kraus}{2013}]{duc13} Duch\^ene G., Kraus A.~L. 2013, ARA\&A, 51, 269
\bibitem[\protect\citeauthoryear{Elliott et al.}{2015}]{ell15} Elliott, P., Hu{\'e}lamo, N., Bouy, H., et al.\ 2015, A\&A, 580, A88 
\bibitem[\protect\citeauthoryear{Elliott \& Bayo}{2016}]{ell16} Elliott, P., \& Bayo, A.\ 2016, MNRAS, 459, 4499 
\bibitem[\protect\citeauthoryear{F\~{u}r\'esz et al.}{2008}]{fur08} F\~{u}r\'esz G., Hartmann L.~W., Megeath T., Szentgyorgyi A.~H., Hamden E.~T. 2008, ApJ, 676, 1109
\bibitem[\protect\citeauthoryear{Goodwin et al.}{2004}]{goo04} Goodwin S.~P., Whitworth A.~P., Ward-Thompsopn D. 2004, A\&A, 423, 169
\bibitem[\protect\citeauthoryear{Goodwin et al.}{2007}]{goo07} Goodwin S.~P., Kroupa P., Goodman A., Burkert A. 2007, in {\it Protostars \& Planets V}, Reipurth B., Jewitt D., Keil K. eds., Univ. of Arizona Press, 133 
\bibitem[\protect\citeauthoryear{Hennebelle \& Teyssier}{2008}]{hen08} Hennebelle, P., \& Teyssier, R.\ 2008, A\&A, 477, 25 
\bibitem[\protect\citeauthoryear{Hillenbrand}{1997}]{hil97} Hillenbrand L.~A. 1997, AJ, 113, 173
\bibitem[\protect\citeauthoryear{Hillenbrand et al.}{1998}]{hil98} Hillenbrand L.~A. 1998, AJ, 116, 1816
\bibitem[\protect\citeauthoryear{Hillenbrand \& Hartmann}{1998}]{hil98b} Hillenbrand, L.~A., \& Hartmann, L.~W.\ 1998, ApJ, 492, 540 
\bibitem[\protect\citeauthoryear{Hillenbrand et al.}{2013}]{hil13} Hillenbrand L.~A., Hoffer A.~S., Herczeg G.~J. 2013, AJ, 146, 85
\bibitem[\protect\citeauthoryear{Hills}{1990}]{hil90} Hills, J.~G.\ 1990, AJ, 99, 979 
\bibitem[\protect\citeauthoryear{Kim et al.}{2016}]{kim16} Kim K.~H. et al. 2016, ApJS, 226, 8
\bibitem[\protect\citeauthoryear{King et al.}{2012}]{kin12} King R.~R., Goodwin S.~P., Parker R.~J., Patience J. 2012, MNRAS, 427, 2636
\bibitem[\protect\citeauthoryear{K\"ohler et al.}{2006}]{koe06} K\"ohler et al. 2006, A\&A, 485, 461
\bibitem[\protect\citeauthoryear{Kounkel et al.}{2016}]{kou16} Kounkel M., Megeath S.~T., Poteet C.~A., Fischer W.~J., Hartmann L. 2016, ApJ, 821, 52
\bibitem[\protect\citeauthoryear{Kounkel et al.}{2017}]{kou17} Kounkel M. et al. 2017, ApJ, 834, 142
\bibitem[\protect\citeauthoryear{Kraus et al.}{2008}]{kra08} Kraus A.~L., Ireland M.~J., Martinache F., Lloyd J.~P. 2008, ApJ, 679, 762
\bibitem[\protect\citeauthoryear{Kraus \& Hillenbrand}{2009}]{kra09} Kraus A.~L., Hillenbrand L.~A. 2009, ApJ, 703, 1511
\bibitem[\protect\citeauthoryear{Kraus et al.}{2011}]{kra11} Kraus A.~L., Ireland M.~J., Martinache F., Hillenbrand L.~A. 2011, ApJ, 731, 8
\bibitem[\protect\citeauthoryear{Kraus et al.}{2012}]{kra12} Kraus A.~L., Ireland M.~J., Hillenbrand L.~A., Martinache F. 2012, ApJ, 745, 19
\bibitem[\protect\citeauthoryear{Kroupa}{1995}]{kro95} Kroupa P. 1995, MNRAS, 277, 1491
\bibitem[\protect\citeauthoryear{Kroupa et al.}{1999}]{kro99} Kroupa P., Petr M.~G., McCaughrean M.~J. 1999, New A., 4, 495
\bibitem[\protect\citeauthoryear{Kroupa et al.}{2001}]{kro01} Kroupa P., Aarseth S., Hurley J. 2001, MNRAS, 321, 699
\bibitem[\protect\citeauthoryear{Kroupa \& Bouvier}{2003}]{kro03} Kroupa P., Bouvier J. 2003, MNRAS, 346, 343
\bibitem[\protect\citeauthoryear{Kroupa et al.}{2018}]{kro18} Kroupa P., Jerabkova T., Dinnbier F., Beccari G., \& Yan Z. 2018, A\&A, in press (arxiv:1801.03095)
\bibitem[\protect\citeauthoryear{Kruskal}{1956}]{kru56} Kruskal J. B. 1956, {\sl Proc. American Mathematical Society}, 7
\bibitem[\protect\citeauthoryear{Lacour et al.}{2011}]{lac11} Lacour S. et al. 2011, A\&A, 532, A72
\bibitem[\protect\citeauthoryear{Lada et al.}{2000}]{lad00} Lada C.~J. et al. 2000, AJ, 120, 3162
\bibitem[\protect\citeauthoryear{Larson}{1969}]{lar69} Larson, R.~B.\ 1969, MNRAS, 145, 271 
\bibitem[\protect\citeauthoryear{Lomax et al.}{2015}]{lom15} Lomax, O., Whitworth, A.~P., Hubber, D.~A., Stamatellos, D., \& Walch, S.\ 2015, \mnras, 447, 1550 
\bibitem[\protect\citeauthoryear{Manara et al.}{2012}]{man12} Manara C.~F. et al. 2012, ApJ, 755, 154
\bibitem[\protect\citeauthoryear{Mardling \& Aarseth}{1999}]{mar99} Mardling R., Aarseth S. 1999, in {\it The Dynamics of Small Bodies in the Solar Systems}, Steves B.~A. Roy A.~E. eds., Kluwer Pub., 385
\bibitem[\protect\citeauthoryear{Marks \& Kroupa}{2011}]{mar11} Marks M., Kroupa P. 2011, MNRAS, 417, 1702
\bibitem[\protect\citeauthoryear{Marks et al.}{2014}]{mar14} Marks M., Leigh N., Giersz M., Pfalzner S., Pflamm-Altenburg J., Oh S. 2014, MNRAS, 441, 3503
\bibitem[\protect\citeauthoryear{Masunaga \& Inutsuka}{2000}]{mas00} Masunaga, H., \& Inutsuka, S.-i.\ 2000, ApJ, 531, 350 
\bibitem[\protect\citeauthoryear{Megeath et al.}{2012}]{meg12} Megeath S.~T. 2012, AJ, 144, 192
\bibitem[\protect\citeauthoryear{Moeckel \& Bate}{2010}]{moe10} Moeckel N., Bate M.~R. 2010, MNRAS, 404, 721
\bibitem[\protect\citeauthoryear{Offner et al.}{2009}]{off09} Offner, S.~S.~R., Klein, R.~I., McKee, C.~F., \& Krumholz, M.~R.\ 2009, ApJ, 703, 131 
\bibitem[\protect\citeauthoryear{Padgett et al.}{1997}]{pad97} Padgett D.~L., Strom S.~E., Ghez A. 1997, ApJ, 477, 705
\bibitem[\protect\citeauthoryear{Parker et al.}{2009}]{par09} Parker R.~J., Goodwin S.~P., Kroupa P., Kouwenhoven M.~B.N. 2009, MNRAS, 397, 1577
\bibitem[\protect\citeauthoryear{Parker \& Goodwin}{2012}]{par12} Parker, R.~J., \& Goodwin, S.~P.\ 2012, MNRAS, 424, 272 
\bibitem[\protect\citeauthoryear{Parker et al.}{2014}]{par14} Parker R.~J., Wright N.~J., Goodwin S.~P., Meyer M.~R. 2014, MNRAS, 438, 620
\bibitem[\protect\citeauthoryear{Patience et al.}{1998}]{pat98} Patience J., Ghez A.~M., Reid I.~N., Weinberger A.~J., Matthews K. 1998, AJ, 115, 1972
\bibitem[\protect\citeauthoryear{Patience et al.}{2002}]{pat02} Patience J., Ghez A.~M., Reid I.~N., Matthews K. 2002, AJ, 123, 1602
\bibitem[\protect\citeauthoryear{Petr et al.}{1998}]{pet98} Petr M.~G., Coud\'e du Foresto V., Beckwith S.~V.~W., Richichi A., McCaughrean M.~J. 1998, ApJ, 500, 825
\bibitem[\protect\citeauthoryear{Price \& Bate}{2008}]{pri08} Price, D.~J., \& Bate, M.~R.\ 2008, MNRAS, 385, 1820 
\bibitem[\protect\citeauthoryear{Raghavan et al.}{2010}]{rag10} Raghavan D. et al. 2010, ApJS, 190, 1
\bibitem[\protect\citeauthoryear{Reipurth et al.}{2007}]{rei07} Reipurth B., Guimar\~aes M.~M., Connelley M.~S., Bally J. 2007, AJ, 134, 2272
\bibitem[\protect\citeauthoryear{Reipurth \& Mikkola}{2012}]{rei12} Reipurth, B., \& Mikkola, S.\ 2012, Nature, 492, 221 
\bibitem[\protect\citeauthoryear{Rizzuto et al.}{2013}]{riz13} Rizzuto, A.~C., Ireland, M.~J., Robertson, J.~G., et al.\ 2013, MNRAS, 436, 1694 
\bibitem[\protect\citeauthoryear{Robberto et al.}{2013}]{rob13} Robberto M. et al. 2013, ApJS, 207, 10
\bibitem[\protect\citeauthoryear{De Rosa et al.}{2014}]{der14} De Rosa, R.~J., Patience, J., Wilson, P.~A., et al.\ 2014, MNRAS, 437, 1216 
\bibitem[\protect\citeauthoryear{Scally et al.}{1999}]{sca99} Scally A., Clarke C., McCaughrean M.~J. 1999, MNRAS, 306, 253
\bibitem[\protect\citeauthoryear{Sicilia-Aguilar et al.}{2005}]{sic05} Sicilia-Aguilar A. et al. 2005, AJ, 129, 363
\bibitem[\protect\citeauthoryear{Siess et al.}{2000}]{sie00} Siess L., Dufour E., Forestini M. 2000, A\&A, 358, 593
\bibitem[\protect\citeauthoryear{Sterzik et al.}{2003}]{ste03} Sterzik M.~F., Durisen R.~H., Zinnecker H. 2003, A\&A, 411, 91
\bibitem[\protect\citeauthoryear{Szegedi-Elek et al.}{2013}]{sze13} Szegedi-Elek E., Kun M., Reipurth B., P\'al A., Bal\'azs L.~G., Willman M. 2013, ApJS, 208, 28
\bibitem[\protect\citeauthoryear{Tobin et al.}{2009}]{tob09} Tobin J.~T., Hartmann L., F\~ur\'esz G., Mateo M., Megeath S.~T. 2009, ApJ, 697, 1103
\bibitem[\protect\citeauthoryear{Tuthill et al.}{2010}]{tut10} Tuthill P. et al. 2010, Proc. SPIE, 7735, 1O
\bibitem[\protect\citeauthoryear{Ward \& Kruijssen}{2018}]{war18} Ward, J.~L., \& Kruijssen, J.~M.~D.\ 2018, MNRAS, 475, 5659 
\bibitem[\protect\citeauthoryear{Ward-Duong et al.}{2015}]{war15} Ward-Duong K. et al. 2015, MNRAS, 449, 2618
\bibitem[\protect\citeauthoryear{Ward-Thompson et al.}{2007}]{war07} Ward-Thompson, D., Andr{\'e}, P., Crutcher, R., et al.\ 2007,in {\it Protostars \& Planets V}, Reipurth B., Jewitt D., Keil K. eds., Univ. of Arizona Press, 33 
\bibitem[\protect\citeauthoryear{Weinberg et al.}{1987}]{wei87} Weinberg, M.~D., Shapiro, S.~L., \& Wasserman, I.\ 1987, ApJ, 312, 367 
\end{thebibliography}








\bsp	
\label{lastpage}
\end{document}